\documentclass{jaa}
\usepackage{graphicx}

\def\approxgt{\mathrel{\hbox{\rlap{\lower.55ex \hbox {$\sim$}}
        \kern-.3em \raise.4ex \hbox{$>$}}}}
\def\approxlt{\mathrel{\hbox{\rlap{\lower.55ex \hbox {$\sim$}}
        \kern-.3em \raise.4ex \hbox{$<$}}}}
\def \xmm {\emph{XMM-Newton}}
\def\pdot {\dot P}

\def\ltsima{$\; \buildrel < \over \sim \;$}
\def\lsim{\lower.5ex\hbox{\ltsima}}
\def\gtsima{$\; \buildrel > \over \sim \;$}
\def\gsim{\lower.5ex\hbox{\gtsima}}

\def\psr {PSR B0943+10}
\def\src {PSR B0943+10}
\newcommand{\aap}{{Astron. Astrophys.}}
\newcommand{\apj}{{Astrophys. J.}}
\newcommand{\apjl}{{Astrophys. J. Lett.}}
\newcommand{\apjs}{{Astrophys. J. Suppl. Ser.}}
\newcommand{\sovast}{{Soviet Astronomy}}

\newcommand{\mnras}{{MNRAS}}

\begin{document}

\title{The radio and X-ray mode-switching pulsar \psr\  }


\author{Sandro Mereghetti\textsuperscript{1,*} \and Michela Rigoselli\textsuperscript{1,2}}
\affilOne{\textsuperscript{1}IASF-Milano, INAF, v. Bassini 15, 20133 Milano, Italy\\}
\affilTwo{\textsuperscript{2}Universit\`a degli Studi di Milano Bicocca, Dipartimento di Fisica, Piazza della Scienza 3, 20126 Milano, Italy.}


\twocolumn[{

\maketitle

\corres{sandro@iasf-milano.inaf.it}

\msinfo{ }{ }{ }

\begin{abstract}
Observations obtained in the last  years challenged the widespread notion that rotation-powered neutron stars are steady X-ray emitters.  Besides a few allegedly rotation-powered neutron stars that  showed  ''magnetar-like'' variability, a particularly interesting case is that of \psr . Recent observations have shown that this pulsar, well studied in the radio band where it alternates between a bright and a quiescent mode, displays significant X-ray variations, anticorrelated in flux with the radio emission.  The study of such synchronous radio/X-ray mode switching opens a new window to investigate the processes responsible for the pulsar radio and high-energy emission.  
Here we review the main   X-ray properties of \psr\ derived from recent  coordinated X-ray and radio observations. 
\end{abstract}


}]


\doinum{12.3456/s78910-011-012-3}
\artcitid{\#\#\#\#}
\volnum{123}
\year{2016}
\pgrange{23--25}
\setcounter{page}{23}
\lp{25}

\section{Introduction}

The   discovery of  X-ray flux variations in \psr\ \cite{her13}, challenged the common view,   derived from decades of observations,  that  rotation-powered neutron stars  are constant X-ray sources.
This unexpected finding has made \src\  a key target for observations which,  by exploiting the new diagnostic tool of correlated radio/X-ray variability, can give some relevant insight into the emission processes at work in radio pulsars. 

In the radio band, \src\ is one of the best studied  pulsars showing  the mode-switching phenomenon. Mode-switching pulsars alternate between two (or more) modes  characterized by different properties of their radio emission, such as, e.g., average pulse profile, intensity, polarization, drift rate of subpulses, etc... \cite{cor13}. 
In fact, at irregular intervals every few hours or less, \src\ switches between a mode  in which  its radio emission displays a regularly organized pattern of drifting subpulses and a mode in which the radio pulses are on average fainter and have a chaotic substructure  \cite{sul84}.

In the 0.2-10 keV energy range  \psr\  is among the  faintest rotation-powered X-ray pulsars \cite{zha05}. Its  luminosity of a few  10$^{29}$ erg s$^{-1}$ corresponds to a $\sim5\times10^{-3}$ efficiency of conversion of   rotational energy   into X-rays.
While this is by no means surprising for a relatively old pulsar with the characteristics of \psr , its  X-ray variability,  correlated with the mode-switching in the radio band, is  remarkably unique.

In this paper, after a short summary of the main radio properties of \psr , we review the  results obtained in two long campaigns of coordinated radio/X-ray observations carried out in 2011 and 2014 and outline a few open questions to be investigated with future observations. 

\section{Main pulsar properties}
\label{sec-main}

\psr\  has a spin period $P$ = 1.1 s and  a period derivative $\pdot$ = 3.5 $\times$ 10$^{-15}$ s s$^{-1}$. These values give a characteristic age $\tau$ = 5 Myr and a  rotational energy loss rate $\dot{E}_{rot}$ = 10$^{32}$ erg s$^{-1}$.
With  the usual dipole  braking assumptions, the corresponding magnetic field at the pole is $B$ = 4 $\times$ 10$^{12}$ G.

For the pulsar dispersion measure of 15.32 pc cm$^{-3}$ \cite{bkk+15}, the Cordes \& Lazio (2002)  model for the Galactic electron density distribution \cite{cor02} gives a  distance $d$ = 0.63 $\pm$ 0.10 kpc. This value has been adopted   in most studies of \psr .  A more recent electron density  model \cite{yao17}  gives instead $d$ = 0.89  kpc.
 
The  Galactic coordinates of \psr\ ($l$ = 225.4$^\circ$,  $b$ = +43.1$^\circ$)  place it  at a height of  0.43$d_{0.63~kpc}$ kpc above the Galactic plane.  The total Galactic column density in this direction is $N_H$ = $2.3\times10^{20}$ cm$^{-2}$ \cite{kal05}, which should represent an upper limit on the absorption of \psr .  However, based on the pulsar dispersion measure, and assuming a 10\% ionization of the interstellar medium \cite{he13}, a value of  $N_H$ = $4.3\times10^{20}$ cm$^{-2}$ has usually been adopted to fit its X-ray spectrum.

With a flux density reaching  $\sim$1 Jy at 100 MHz,  \src\ is one of the brightest pulsars at low frequency \cite{bkk+15}.  Its 0.1-10 GHz   spectrum is very steep (S$_{\nu}\propto \nu^{-2.9}$, \cite{mal00}). This is probably due to  the observing geometry: the line of sight is grazing a conal beam which becomes narrower with increasing frequency. 

Early estimates of the geometry and orientation of \src\ were based on the assumption of a conal beam geometry, supported by the data on radio profiles and polarization, and indicated a relatively small angle  between the rotation and magnetic axis $\alpha\sim11-12^{\circ}$ \cite{lyn88,ran93b}.

One expects to see the phenomenon of drifting subpulses  when the line of sight crosses nearly tangentially the hollow cone of emission which characterizes the radio beam of a pulsar.  In this framework,  a  determination of  the geometry and orientation of \psr\  was derived thanks to  a detailed modeling of the average pulse profile and of the drifting subpulses  \cite{des01}.  These authors  estimated  $\alpha$ based on different plausible assumptions on the width of the emission cone and its frequency dependence.
In particular, they derived $\alpha=11.6^{\circ}$ or $\alpha=15.4^\circ$, in case we are seeing a so called  ``inner'' or  ``outer'' cone, respectively.  
The corresponding angles between the line of sight and the rotation axis are  $\xi=7.3^\circ$ or  $\xi=9.7^\circ$. This configuration, with the line of sight between the rotation and magnetic axis, is commonly denoted ``inside traverse''.   It should be kept in mind that such estimates do not imply that the angles are known with an accuracy at degree level, since they  are clearly model dependent and subject to systematic uncertainties of fits with with many parameters which are difficult to quantify (see, e.g. \cite{eve01}). However, as also shown by the more recent analysis of \cite{bil14}, it seems reasonable to assume that \psr\ is a nearly aligned rotator seen approximatley pole-on.
The fact that the  thermal  X-ray  emission is modulated at the spin period and shows a single broad pulse  aligned in phase with the main radio pulse  (see Section~\ref{sec-X}), is also consistent with the hypothesis that  we are seeing emission from a  single region at (or close to) the magnetic polar cap.

\begin{figure}[h]
\hspace{-1cm}
\includegraphics[angle=-90,width=10cm]{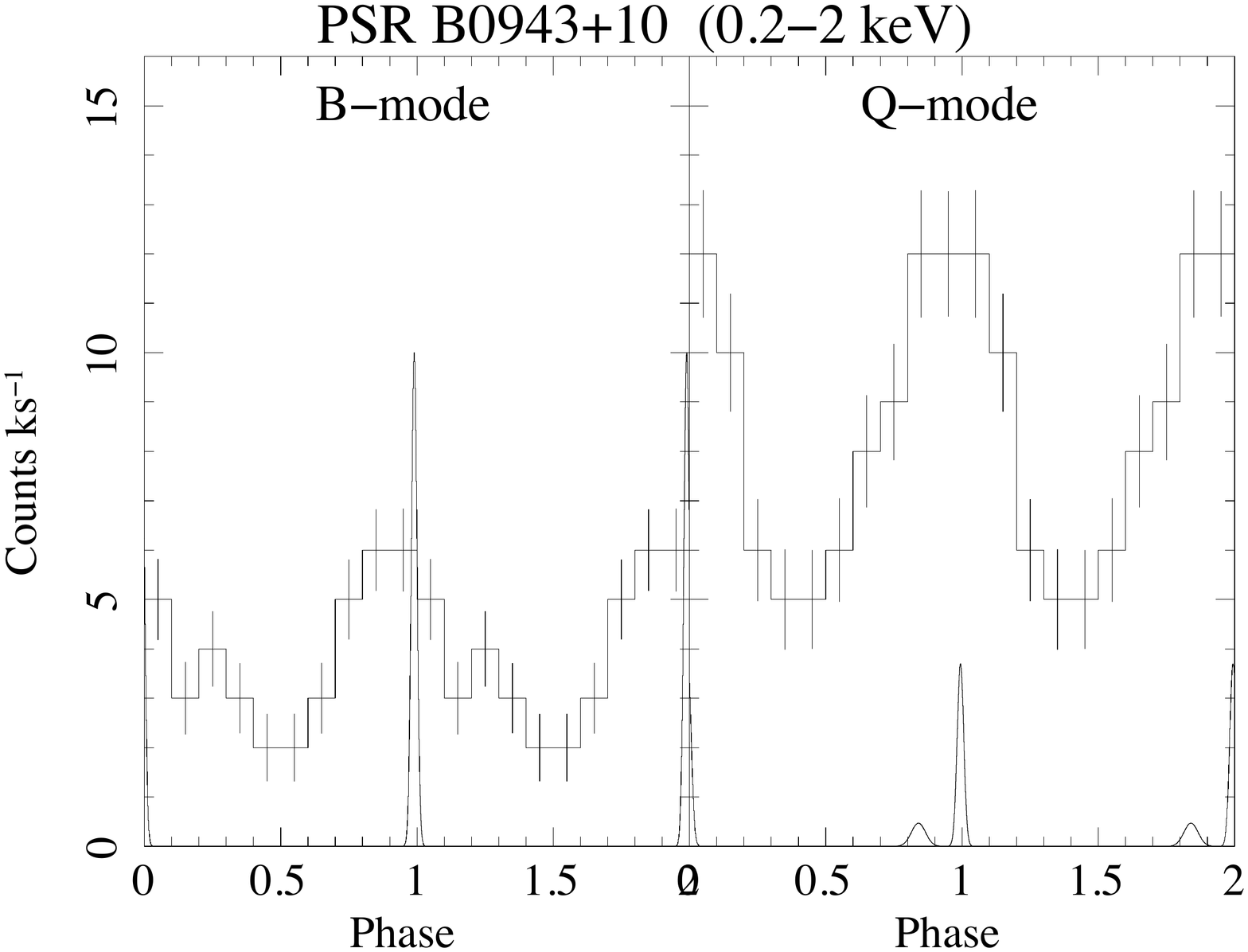}
\caption{\label{fig_lc}  The histograms are the background-subtracted light curves of \psr\ in the 0.2-2 keV energy range obtained in the 2014 \xmm\ observations. The pulsed fractions, derived by fitting sinusoidal profiles, are (43$\pm$7)\% and (42$\pm$5)\% in the B- and Q-mode, respectively. The radio profiles (at  0.3 GHz, in arbitrary intensity units, \cite{her13}) are also indicated.}
\end{figure}

\section{Radio modes}
\label{sec-modes}
        
The bimodal behavior of \psr\ was first noticed at low frequency (62 MHz) and then confirmed  by  observations at 102 MHz with the Pushchino radiotelescope  \cite{sul84}. The two modes were originally named  ``quiescent'' and ``bursting'' (Q- and B-mode for short). The latter is also  commonly called  ``bright mode'',  although only   the radio emission is brighter:  indeed, the X-rays  in the B-mode are   a factor \gtsima2 {\it dimmer} than in the Q-mode.      

The average   pulse profile in the B-mode consists of two nearly Gaussian components, with a  separation which decreases with increasing frequency  from $\sim20^\circ$ at 26 MHz until they merge into a single peak of width $\sim15^\circ$ above 180 MHz \cite{bil14}. This is  consistent with a hollow-cone radio beam geometry, where emission of lower frequencies is produced at larger heights in the magnetosphere \cite{cor78}.  

The most striking characteristic of the B-mode is the presence of very regular drifting subpulses. This phenomenon  gives rise to a series of observed periodicities, besides that associated with the neutron star spin at $P_1$ = 1.098 s. They include  a strong and very stable modulation  at frequency  $f_p$ = 0.465 cycles/$P_1$  (=  0.42  Hz,   2.36 s) and a weaker one at $f_s$ $\sim$ 0.07 cycles/$P_1$  (= 0.064 Hz, 15.7 s).

During the Q-mode, the regular pattern of drifting subpulses disappears, the mean pulse profile is weaker and broader,  the radio subpulses have a chaotic modulation pattern and are on average fainter  (altough some of them can be individually brighter than the B-mode ones) \cite{sul98,ran03}. A further difference is the presence, only in the Q-mode, of a strongly linearly-polarized precursor pulse of radio emission centered 52$^\circ$ before the main pulse \cite{des01,bac10}.

There is no consensus in the pulsar community on what causes the phenomenon of drifting subpulses, which is observed in a large number of objects. A variety of models have been proposed based on very different scenarios including oscillating standing waves, emission regions rotating around the magnetic or the spin axis, plasma instabilities, neutron star oscillations (see, e.g., \cite{kui09} and references therein). 
In the frame of the ``rotating carousel'' model \cite{des01,gil00}, the drifting subpulses are associated with emission from  20  polar cap sparking regions that rotate around the magnetic axis due to the    $\vec{{\sl E}}\times\vec{{\sl B}}$ drift effect \cite{rud75,gil00}. A complete rotation of the carousel takes $P_4=20/(1-f_{p})$ = 40.7 s   (i.e.  37.4 rotations of the neutron star,  $f_p$ is the alias of the true frequency associated to the longitude interval between subbeams).

It was initially believed that   \psr\ had a similar probability of being in any of the two modes, but it has been shown by more recent, long observations that the B-mode is dominant. The average durations of the B- and Q-modes are about 7.5 and 2.2 hours, respectively \cite{bac11}. The modal  transitions occur very rapidly for what concerns the regular pattern of the subpulses, that appears/disappears in less than one rotation period, while more gradual variations are seen to affect the average pulse profile  \cite{sul98}.  In fact, after the Q-to-B transitions, several properties of the radio emission gradually evolve toward an asymptotic state on a time scale of about one hour:   the ratio of the amplitudes of the trailing and leading components of the pulse decreases,  the linear polarization increases,   the carousel circulation time increases  \cite{ran06,sul09,bac11}.

\begin{figure}[h]
\vspace{-2cm}
\includegraphics[width=1\columnwidth]{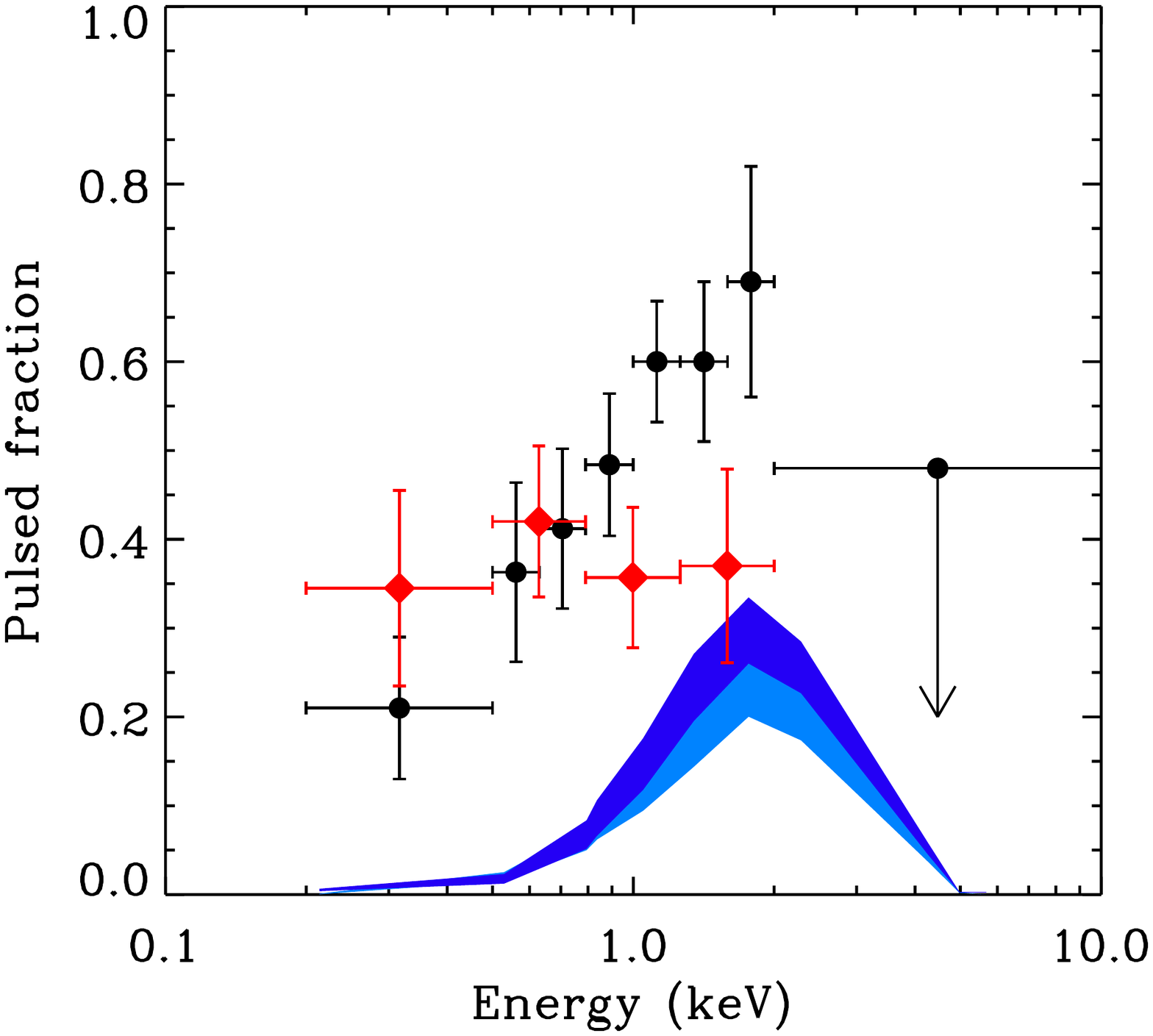}
\vspace{-3cm}
\caption{Pulsed fraction  as a function of energy for the Q-mode (black dots) and for the B-mode (red squares). The blue regions indicate the values expected from the polar cap emission with model atmospheres of different temperatures and using  the $\alpha$  and $\xi$ angles derived for \psr\ (see Section \ref{sec-pt} for details).  }
\label{fig_pf}
\end{figure}

\section{X-ray observations}
\label{sec-X}

\subsection{First observations and discovery of variability}
\label{sec-Xold}

The first X-ray observations of \psr\ were carried out with the  \xmm\ satellite in April and December 2003  and yielded about 30 ks of data, after filtering out high background time periods which affected most of the first observation. The pulsar was significantly detected with a flux of $\sim$ 5 $\times$ 10$^{-15}$ erg cm$^{-2}$ s$^{-1}$, but the low count rate hampered a detailed analysis. In particular, it was not possible to distinguish between thermal and non-thermal emission and  pulsations were not detected \cite{zha05}. Due to the lack of   radio data, the pulsar mode during the 2003 X-ray observations is not known with certainty (our reanalysis of these data indicates that the pulsar was most likely in B-mode during the December 2003 observation). 
                    
In 2011, the first  X-ray  observations with simultaneous radio coverage  (\xmm\ plus GMRT at 320 MHz and LOFAR at 140 MHz)  led to the surprising discovery of X-ray variability anti-correlated with the pulsar radio intensity \cite{her13}. With more than 100 ks of useful data  (nearly equally split between the two modes) it was clearly seen  that the X-ray emission was a factor $\sim$2.5 {\it brighter}  during the {\it radio-fainter} Q-mode.  
X-ray pulsations at the neutron star spin period  were revealed in the Q-mode, but not in the   B-mode. 
Either a blackbody or a power law  could fit equally well the B-mode spectrum. On the contrary, single component models were  ruled out for the Q-mode, which required the sum of a blackbody with temperature kT $\sim$ 0.28  keV and a power law with photon index $\Gamma\sim$ 2.6. The pulsed spectrum (in Q-mode) was found to be a blackbody, consistent in flux and temperature with the blackbody component of the total spectrum.

It was thus concluded that when \psr\ is in B-mode it emits only an unpulsed non-thermal component and the higher luminosity of the Q-mode is caused by the addition of a 100\%-pulsed thermal component \cite{her13}.
Considering the difficulty to produce  such a large modulation in the thermal emission, given  the  likely small values of the angles $\alpha$ and $\xi$,  a different interpretation was put forward by \cite{mer13},  who attributed the flux increase and pulsations in the Q-mode to the addition of a non-thermal component, superimposed to a thermal component present in both modes.

\subsection{The long  campaign of November 2014}
\label{sec-xnew}

To better constrain the X-ray spectral and variability properties of \psr  , the two groups cited above joined their efforts in a long observing campaign carried out in November 2014 \cite{mer16}. The X-ray data were obtained thanks to an \xmm\ Large Program and consisted of seven uninterrupted pointings with durations between 16  and 19  hours each. Simultaneous radio coverage was provided by  the LOFAR, LWA and Arecibo radiotelescopes.  The coordinated use of ground facilities at different longitudes made it possible to obtain a continuous radio coverage, as well as  to observe a few complete instances of  both modes.  These observations yielded  useful X-ray esposures of $\sim$120 ks  and $\sim$175 ks  for the Q- and B-modes, respectively:  an increase by  factors 2.4  and  3.5 compared to the previous data.  

A first important result, obtained thanks to the improved counts statistics, was the discovery that the X-ray flux is modulated at the pulsar spin period also during the  B-mode (see Fig.~\ref{fig_lc}). The pulsed fraction is about 40\%  in the 0.2-2 keV range. Contrary to the Q-mode, in which the pulsed fraction shows a significant energy-dependence,  there is no evidence, within relatively large errors, that the pulsed fraction in B-mode  depend on energy (see Fig.~\ref{fig_pf}).

The new data allowed also a much better characterization of the B-mode spectrum, that, at variance with the findings of the 2011 observations, could not be fit by a single power law, thus ruling out the interpretation of \cite{her13}. Either a single blackbody, with temperature $kT\sim$0.23 keV,  or a blackbody plus power law, with parameters similar to those of the Q-mode, provide a good fit to the B-mode spectrum \cite{mer16}.

\begin{figure*}[ht!]
\includegraphics[angle=0,width=18cm]{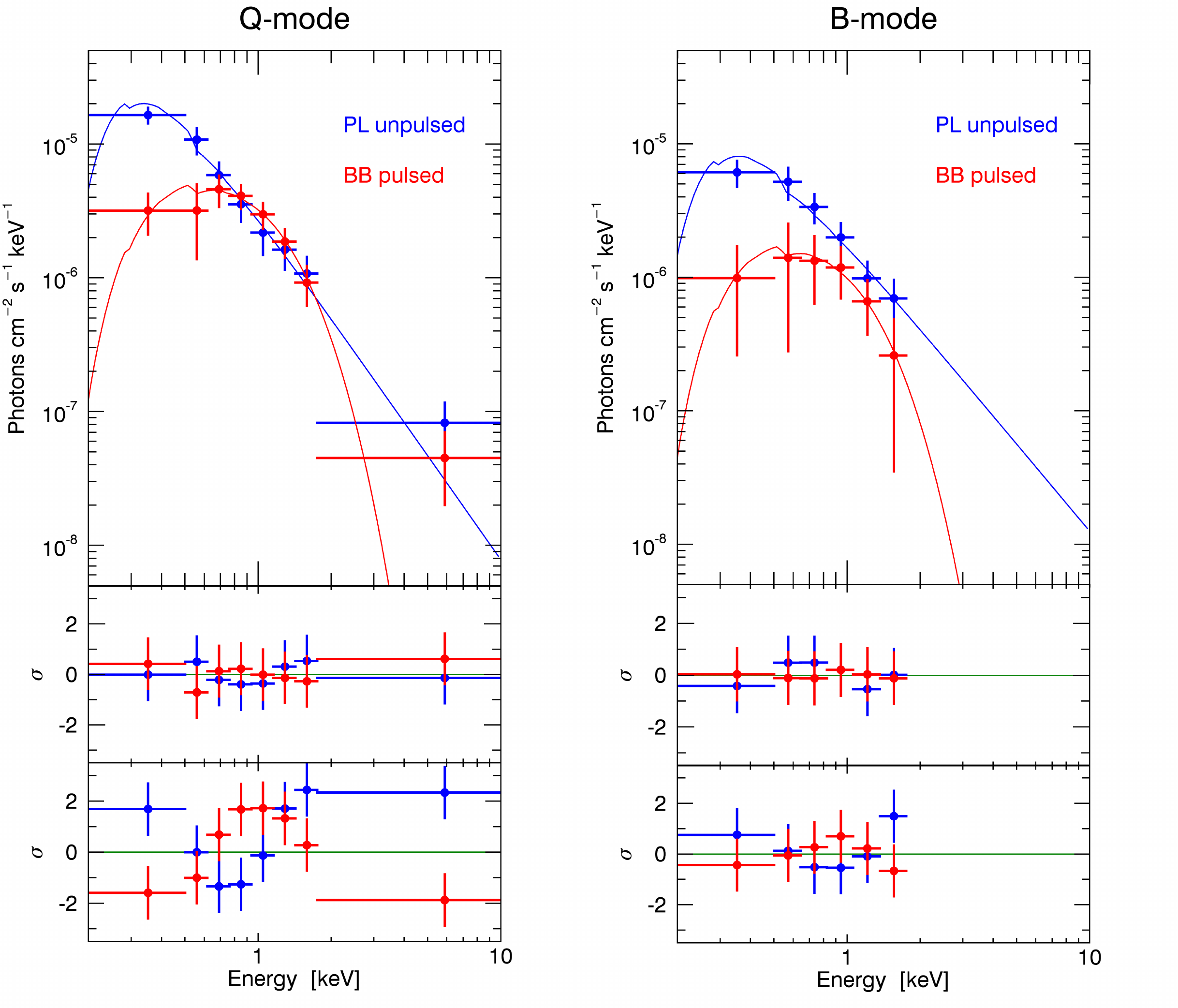}
\caption{\label{fig_spectra} X-ray spectra of the unpulsed (blue) and pulsed (red) emission from  \psr\ in the Q-mode (left panels) and B-mode (right panels), obtained with the EPIC pn instrument.  The top panels show the best fits in which the unpulsed flux is modeled with a power law and the pulsed flux with a blackbody. The corresponding residuals are shown in the middle panels.  The bottom panels show instead the residuals obtained by fitting the unpulsed flux with a blackbody and the pulsed flux with a power law. Interstellar absorption with an equivalent column density fixed at N$_H$ = 4.3$\times$10$^{20}$ cm$^{-2}$ has been included in all the fits.}
\vspace{0.5cm}
\end{figure*}

Very interesting results were obtained by applying a 3-D maximum likelihood analysis \cite{her17} to {\it simultaneously} derive the spectra of the pulsed and unpulsed emission.  
The  findings of \cite{her13} for the Q-mode were confirmed  (Fig.~\ref{fig_spectra}, left panels):  the pulsed flux is thermal, being well fit by a blackbody but not by a power-law. The opposite is true for the unpulsed flux, which can be fit by a power-law and not by a blackbody.  The blackbody describing  the pulsed emission and the power law describing the unpulsed emission are consistent with the two components used to fit the total spectrum of the Q-mode. 
The situation is less well constrained for the B-mode, where, both for the pulsed and unpulsed emission, it was impossible to distinguish between a  power law and a blackbody which give similarly good fits (Fig.~\ref{fig_spectra}, right panels).

\subsection{A scenario for the X-ray variability}
\label{sec-scenario} 

After the 2011  observations of \psr , the apparent lack of pulsations, lower flux and simpler spectrum of the B-mode prompted attempts to interpret its  X-ray variability   in terms of a minimal set of emission components: a  steady one, present with the same properties in both modes, and an additional pulsed component, appearing only in the Q-mode, causing  the higher flux \cite{her13,mer13}.  
These interpretations are appealing for their simplicity. However, they are challenged by some difficulties, especially if the variability is attributed to the thermal component. Besides  being able to  appear and disappear very rapidly at the modal transitions, such a thermal component should be  $\sim$100\% pulsed, which seems difficult to reconcile with the pulsar geometry and pole-on line of sight described in Section \ref{sec-main}
Indeed a more complex picture emerged from the   data of the longer 2014 campaign, which showed that the properties of the X-ray emission in the two modes are more similar than previously thought. 

In view of the two-components spectrum seen in the Q-mode, it is quite reasonable to assume the simultaneous presence of thermal and non-thermal X-rays also during the B-mode \cite{mer16}. 
In fact, although a single blackbody gives by itself a formally acceptable fit to the B-mode spectrum, this can also be fit by the sum of  a blackbody plus  a power law. 
When the X-ray luminosity is  higher (Q-mode),  the two components have approximately the same flux\footnote{all the quoted fluxes are corrected for the interstellar absorption (4.3$\times$10$^{20}$ cm$^{-2}$) and refer to the 0.5-2 keV energy range.} of 6$\times$10$^{-15}$ erg cm$^{-2}$ s$^{-1}$. When the pulsar switches to the  B-mode, the thermal flux decreases by a factor $\sim$3 and the non-thermal one decreases by a factor $\sim$2. Based on  the  current data,  the pulsations can be entirely attributed to the thermal component in both modes.

In the following we  concentrate on this scenario, but it should be  kept in mind that   even more complex situations cannot be excluded  (e.g. both the thermal and non-thermal emission could be pulsed, additional components might be present, etc...), as might be revealed by future X-ray observations of   better statistical quality.

\section{Origin of the X-ray emission}

\subsection{The unpulsed non-thermal component}

The non-thermal emission from \psr\ is well represented by a power-law with photon index $\Gamma\sim2.5$ in both radio modes  and  with  luminosities   $L_B$ = 3.8 $\times$ 10$^{29}$ erg s$^{-1}$  and $L_Q$ = 8.1 $\times$ 10$^{29}$ erg s$^{-1}$ (unabsorbed, 0.2-10 keV,  $d$ = 0.63 kpc).
The corresponding efficiency values of about 0.004 and 0.008 are in the typical range observed in radio pulsars. 
The  variability excludes the presence of a  major contribution from a pulsar wind nebula. Indeed there is no evidence for diffuse X-ray emission surrounding \psr\ in \xmm\ or  in {\it Chandra} data  \cite{mer16},  as expected for a relatively old pulsar.

Non-thermal emission in rotation-powered pulsars is usually attributed to synchrotron,  inverse Compton, and curvature radiation from electrons and positrons accelerated in the magnetosphere. Differerent locations of the particle acceleration sites have been considered, ranging from regions close to the polar caps \cite{har09}, to ``outer gaps'' accelerators close to the light-cylinder \cite{che09},  and even farther out with models of emission in the striped wind \cite{pet11}.  The role of backward-accelerated particles in heating the star surface has been explored in details by \cite{har01,har02} in the context of polar cap models.  The observation of  a correlation between the intensity of the thermal and non-thermal X-ray components in \psr\ supports a causal relation between the flow of accelerated particles and polar cap heating,  although the quantitative details of this correlation remain to be explained. It is clear that geometrical effects also play an important role in determining what we actually can observe, thus contributing to the different ratios of thermal versus non-thermal X-rays seen in rotation-powered neutron stars.  Future X-ray observations of \psr\ with more sensitive instruments will offer the advantage of studying this relation in a single object, i.e. at a fixed geometry.

\subsection{The pulsed thermal component}
\label{sec-pt}

It is natural to associate the blackbody-like component with thermal emission from  the star surface. Given the relatively old characteristic age of \psr , we do not expect detectable X-ray emission from the bulk of its surface, which should have cooled by now to temperatures well below 10$^6$ K.  Indeed the observed blackbody flux requires emission only from a small area.
\psr\ is thus similar to other old and middle aged pulsars from which thermal X-rays from a small emitting area have been detected (e.g. \cite{mis08,gil08,pos12,del05}). It is   believed that the surface heating in these pulsars results from the flow of backward-accelerated particles produced in the star magnetosphere above the polar caps \cite{har01,gil07}. 
 
The radius of the emitting area inferred from the blackbody fits is of the order of  $R_{\rm th}\sim$20 $\left(\frac{d}{630~\rm{pc}}\right)$  m.  This is significantly smaller than the radius of the classical polar cap  defined by the last open dipole field line,  $R_{\rm{PC}}$=$\left(\frac{R_{\rm{NS}}}{R_{\rm{LC}}}\right)^{1/2} R_{\rm{NS}}\sim$140  m  (for $R_{\rm{NS}}$=10 km).
Note, however,  that fits of the thermal emission with  neutron star atmosphere models rather than with a blackbody, yield different values for the temperature and emitting radius.
For example, a partially ionized hydrogen atmosphere with $B$ = 2 $\times$ 10$^{12}$ G gives a temperature   a factor $\sim$2 smaller than the blackbody value and an emitting radius of $\sim$85 m \cite{sto14}, closer to $R_{\rm{PC}}$.
The currently available data do not allow to significantly discriminate between simple blackbody emission and different more complicated atmosphere models.
 
\begin{figure}[h]
\hspace{-1cm}
 \includegraphics[angle=0,width=11cm]{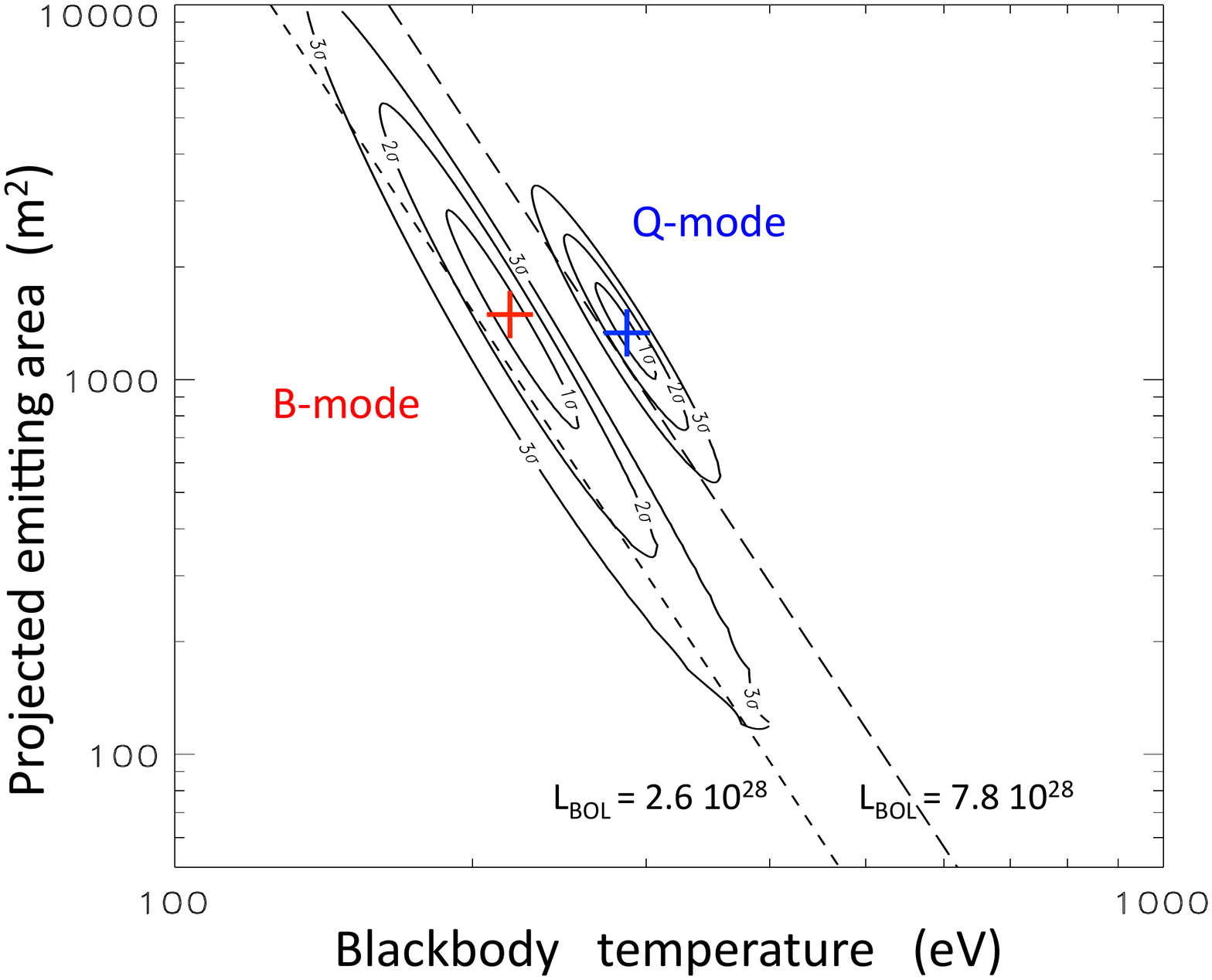}
\caption{\label{fig_cont_kt_area}  Error regions of the parameters derived for the thermal emission of \psr\ in the two modes (adapted from \cite{mer16}). The crosses indicate the best fit values of blackbody temperature and projected emitting area,  while the contours show the  corresponding uncertainty regions  (1, 2, and 3$\sigma$ c.l.). 
The dashed lines indicate values of the  bolometric luminosities (in  erg s$^{-1}$)  of the polar cap, corrected for the (small) projection effect and assuming  $d$=630 pc.}
\end{figure}

An interesting open  question is whether the higher luminosity of the thermal emission in the Q-mode is primarily caused by a change in the temperature or in the area of the emitting region. The best fit blackbody temperatures for the pulsed thermal component in the two modes are  $kT_B$ =  0.22$^{+0.04}_{-0.03}$ keV   and $kT_Q$ = 0.29$\pm$0.02 keV  (\cite{mer16}, 1$\sigma$ errors), while the corresponding emitting radii are consistent with the same value within 1$\sigma$.  The  correlation between the errors in temperature and projected emitting area\footnote{for the geometry of \psr ,  the projected area of a small polar cap is close to the real one (the cosine of the angle between magnetic axis and line of sight, averaged over the pulsar rotation phase is $\sim$0.96).} is illustrated  by the confidence regions plotted in Fig.~\ref{fig_cont_kt_area}. The best fit values suggest that the flux variation is caused by a temperature change, but further observations with higher statistics are needed to confirm this.

Another puzzling result concerning the non-thermal emission is its relatively large pulsed fraction (Fig.~\ref{fig_pf}).   If \psr\ is really  a nearly aligned rotator seen from a direction close to the rotation and to the magnetic axes ( as believed by most authors \cite{lyn88,ran93b,des01,bil14}),  the X-ray emitting region   should be always visible,  if it is located not far from the pole of the dipolar magnetic field. As the pulsar rotates, the orientation of the X-ray emitting hot spot with respect to the line of sight should not vary much, and thus it should produce  a shallow light curve with a small pulsed fraction. 
 
The light blue region in Fig.~\ref{fig_pf} shows the pulsed fraction  expected from a small polar cap with effective temperature $kT$ = 0.25 keV, computed using a  hydrogen atmosphere model with $B=4\times 10^{12}$ G,  and the  geometrical angles in the range of values expected for \psr\ (see details in \cite{mer16} and \cite{tur13}).  Slightly larger pulsed fractions are obtained with $kT$ = 0.14 keV (dark blue region in Fig.~\ref{fig_pf}), but still smaller than the observed values.
The discrepance suggests  that some  of the above assumptions are incorrect.  One could  invoke different magnetic and viewing angles,  beaming effects, or   configurations of the magnetic field more complex  than a  dipole. Indeed,  detailed models of multipolar fields in \psr\ have  been presented by several authors \cite{ass02,sza13} and an offset dipole was recently considered specifically to explain its X-ray pulsed emission \cite{sto14}.  Unfortunately the statistical quality of the current X-ray data is unable to univocally constrain the large number of  parameters involved in such models.

\subsection{Mode transitions and evolution within modes}

\begin{figure}[h]
\hspace{-1cm}
 \includegraphics[angle=-90,width=10cm]{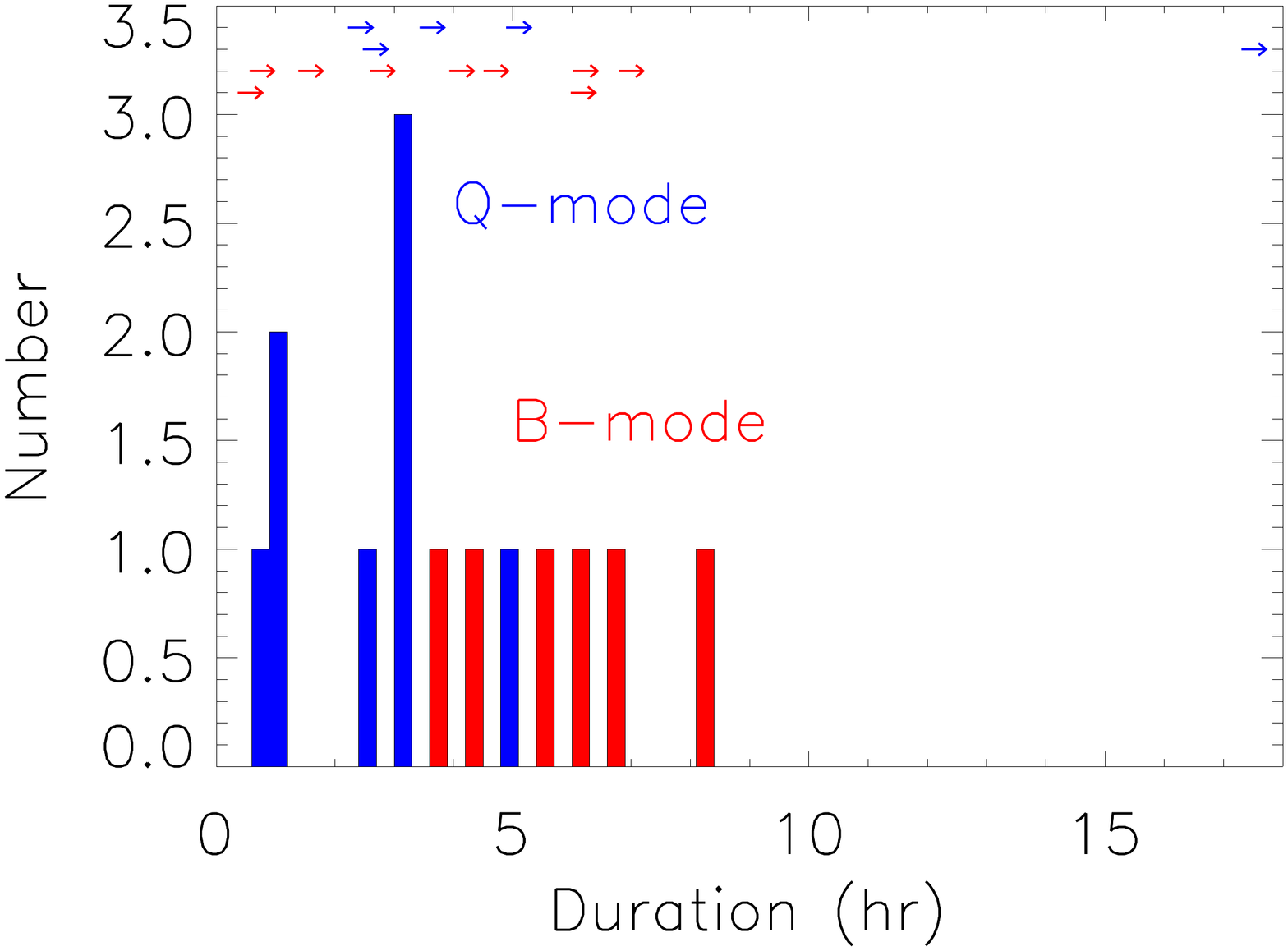}
\caption{\label{fig-dur}  Distributions of the durations of B- (red) and Q-modes (blue) during the simultaneus radio/X-ray observations of November 2014 \cite{mer16}. The arrows in the upper part of the figure indicate lower limits. The B-modes tend to last longer than the Q-modes, but note the occurrence of a very long Q-mode that lasted  more than 17 hours. 
 }
\end{figure}

Thanks to the long uninterrupted X-ray pointings of the  2014 campaign, it was  possible to obtain a few complete instances of each mode and to observe several mode transitions.  The distribution of the mode durations for the 2014 observations is shown in  Fig.~\ref{fig-dur} (8 complete Q-modes and 6 complete   B-modes; in 2011 only one complete B-mode, lasting about 3 hr was seen).

\begin{figure}[h]
\hspace{-1cm}
 \includegraphics[angle=-90,width=10cm]{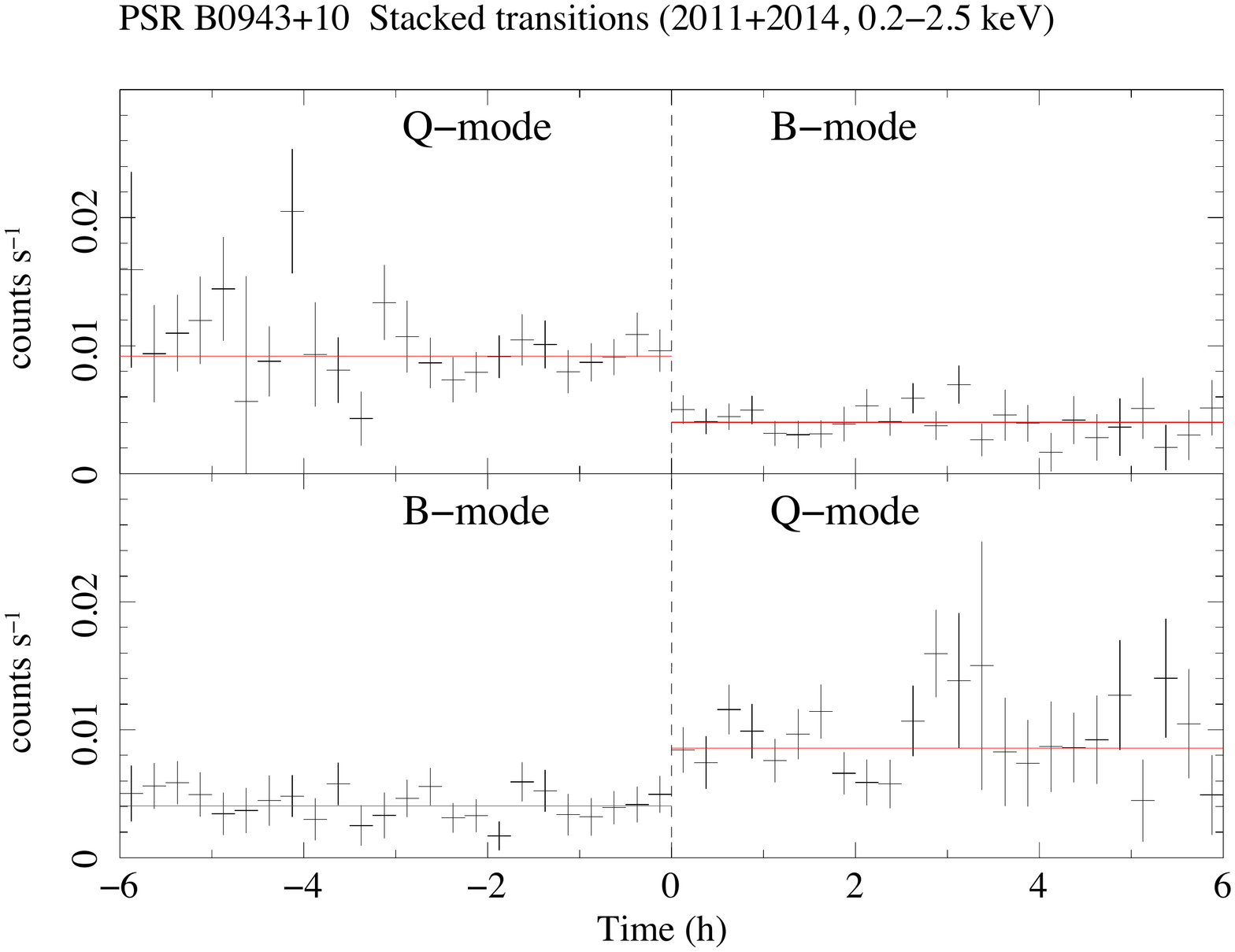}
\caption{\label{fig-stack}  Background-subtracted and exposure-corrected light curves in the 0.2-2.5 keV range obtained by stacking all the Q to B (top panel) and B to Q (bottom panel) mode transitions observed in 2011 and 2014.  The bin size is 15 minutes. 
The red lines are the best fit  constant rates with the following values. 
Q-mode before transition: 0.0092$\pm$0.0004 cts s$^{-1}$, 
B-mode after transition:     0.0040$\pm$0.0003 cts s$^{-1}$, 
B-mode before transition:  0.0041$\pm$0.0003 cts s$^{-1}$, 
Q-mode after transition:    0.0085$\pm$0.0005 cts s$^{-1}$. 
 }
\end{figure}

While mode transitions have been studied with very good time resolution in the radio band, where single pulses can be measured, similar studies at X-ray energies are impossible with the current facilities. Consider that the existing instrument with the largest collecting area in the $\sim$keV region,  EPIC   on \xmm , yields an average rate of only one photon from  \psr\   every $\sim$1.5 minutes (in the Q-mode and summing the three EPIC cameras). The  counting statistics to study the transitions can be increased by stacking the data with  the times properly shifted in order to align the mode  transitions (15 Q-to-B  and  12 B-to-Q transitions have been observed in X-rays up to now). An example of this is illustrated by the light curves plotted in Fig.~\ref{fig-stack},  which have a  bin size of 900 s. They show that  the pulsar X-ray count rates before/after the transitions are consistent with constant values.

The stacked data were also used to search for possible  variations within the modes.   This resulted in  some  evidence  for an increase of the pulsed fraction during the B-mode, which passed from a value  of   27$\pm$8\%   measured during the first three hours after the Q-to-B transitions, to a value of  42$\pm$8\%   measured in the following three hours \cite{mer16}. Although the difference is only at a $\sim2\sigma$ level, this might be an interesting finding worth to be checked with better data, in view of the gradual changes in the the radio properties during the B-mode mentioned in Section~\ref{sec-modes}.   No variations were found in the X-ray pulsed  fraction during the Q-mode.

\begin{figure}[h]
\hspace{-1cm}
 \includegraphics[angle=90,width=10cm]{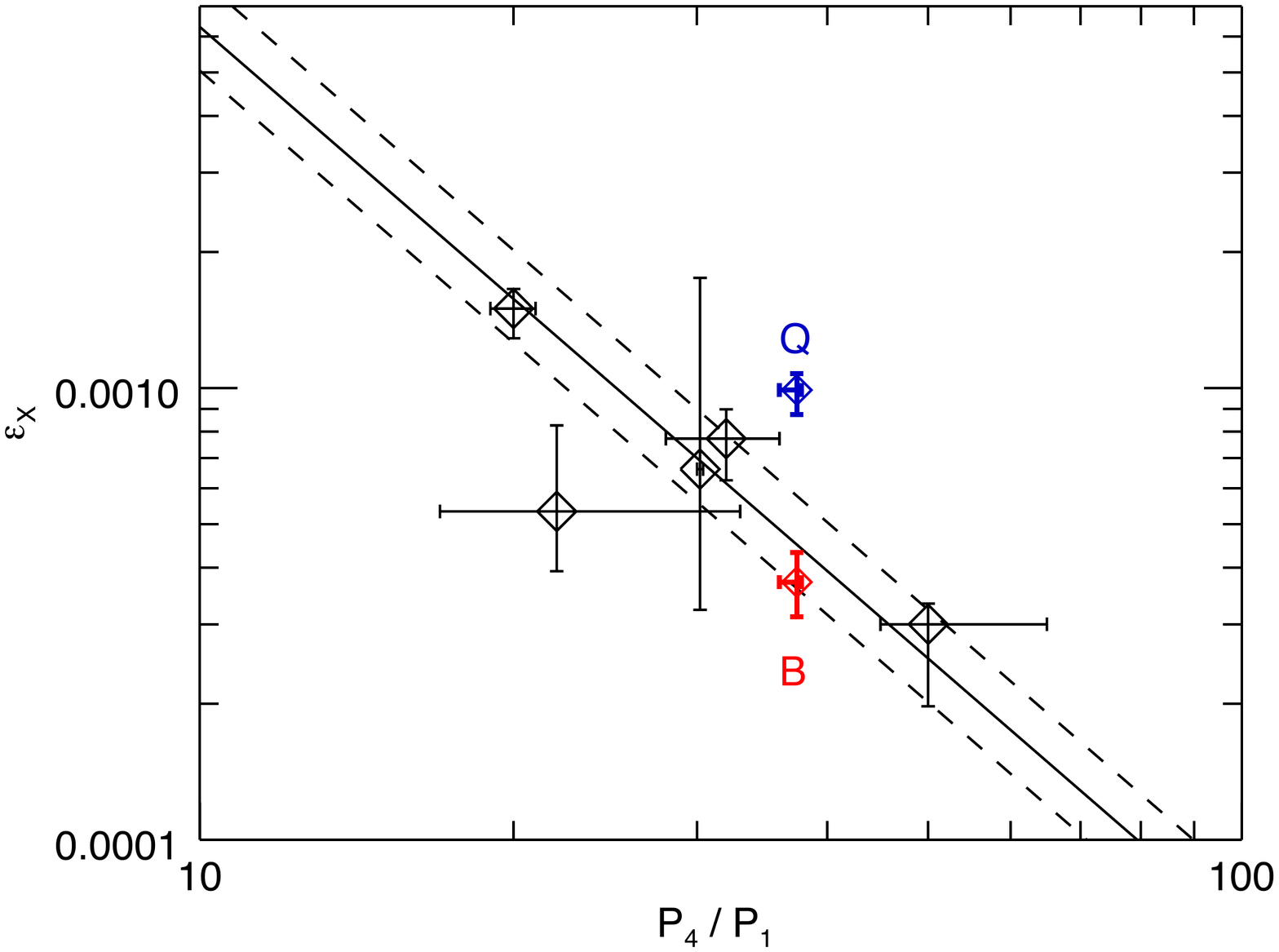}
\caption{\label{fig-psg-1}  Relation between the X-ray efficiency   $\epsilon_X$ = $L_{BOL}$ / $\dot{E}_{rot}$ and the carousel circulation time (in units of spin period)  $P_4/P_1$, as predicted by the PSG model (adapted from \cite{gil08}).
 The points for \psr\ refer to the B-mode (red) and  Q-mode (blue) for  $d$ = 630 pc. The other pulsars plotted in the figure are (from left to right):  B0656$+$14, B1055$-$52, B0834$+$06, B1133$+$16 and B1929$+$10. }
\end{figure}

\section{Discussion}

\subsection{\psr\ in the Partially Screened Gap model}
\label{sec-psg}

In this subsection we discuss the   \psr\ observations in the context of the Partially Screened Gap model, with the caveat that this is only one of the possible interpretations.

The appearance of drifting subpulses in several radio pulsars has been attributed to the ordered motion of  spark filaments of plasma in the inner acceleration region above the polar cap.  Due to the non-corotational value of the charge density, such plasma filaments, together with the related regions higher in the magnetosphere where radio emission is produced,  are subject to an     $\vec{{\sl E}}\times\vec{{\sl B}}$   drift that makes them rotate around the magnetic axis \cite{rud75}. This physical mechanism forms the basis of the partially screened gap (PSG) model, which modifies the original ``pure vacuum gap'' scenario by including the effect of partial screening of the polar cap electric field by ions thermally emitted from the neutron star  surface \cite{gil03,gil06}. 

One of the motivations to introduce the PSG model was that the drift rate predicted by the pure vacuum  model is too fast, compared to that deduced from the analysis of drifting subpulses.  In the PSG model both the circulation period of the spark carousel, $P_4$, and the heating rate of the polar cap depend on the value of the electric field in the gap. This leads to a well defined relation between the circulation period,  the thermal bolometric luminosity and the rotational energy loss rate that can be compared with the observations (see Fig.~\ref{fig-psg-1}).  It can be seen that \psr\  agrees well with the model predictions, especially during the B-mode, i.e. when the rotating carousel is present (note, however, that  for the revised DM distance of 890 pc, the luminosities of \psr\  would be larger than those indicated in the figure  by a factor two). 

\begin{figure}[h]
\hspace{-2cm}
\centering \includegraphics[angle=90,width=10cm]{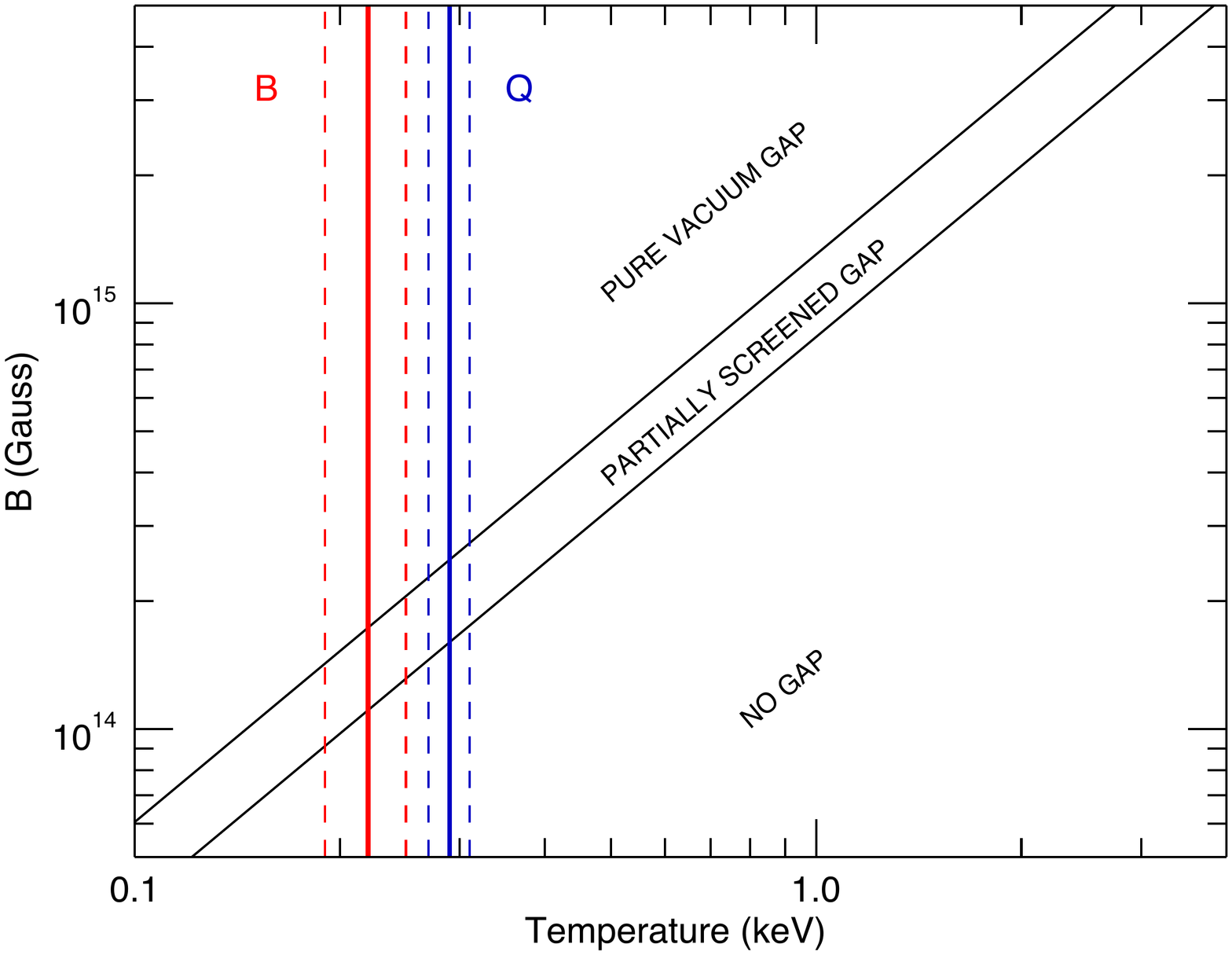}
\caption{\label{fig-psg-2}  Schematic illustration of the different conditions above the polar cap as a function of temperature and magnetic field. The PSG model can only work for temperatures in a critical range approximately indicated by the two diagonal lines \cite{gil06}.
The vertical lines indicate the temperatures (with their 1$\sigma$ uncertainties) for the B-mode (red) and Q-mode (blue) of \psr . They imply surface magnetic field strengths of a few 10$^{14}$ G, well above the dipolar value of 4 $\times$ 10$^{12}$ G.}
\end{figure}

A PSG can form if the temperature and magnetic field conditions at the polar cap  are adequate to partially bind the iron ions to the star surface. If the temperature is too high and/or the magnetic field too low, too much ions are emitted and no vacuum gap can form. In the opposite conditions a pure vacuum gap is formed. This is schematically illustrated in Fig.~\ref{fig-psg-2}.  
Since the surface heating is due to the flow of particles back-accelerated in the gap, a self-regulating thermostatic mechanism acts to keep the temperature at the value required by the PSG model to work \cite{gil03}.  As shown in Fig.~\ref{fig-psg-2}, the   temperatures measured for the polar cap in \psr\ imply  a magnetic field of the order of $\sim10^{14}$ G. This is stronger by at least a factor 25 than  the dipole field inferred from the timing parameters, requiring  the presence of strong magnetic field structures close to the star surface.  The Hall drift of a large scale toroidal magnetic field in the netutron star crust has been suggested as a likely source of these strong multipolar components near the star surface \cite{gep13}.

However, we finally note that the X-ray results obtained for \psr\ cannot either prove or exclude the presence of a non-dipolar field. This can only be done based on model-dependent interpretations of the observations (for example, the size inferred for the thermally emitting area can be reconciled with a dipole polar cap if the spectrum differs from a blackbody and/or a larger distance is assumed). 

\subsection{Global or local magnetospheric changes?}

Although it is widely believed that the study of mode-switching,  and of possibly related phenomena such as nulling, pulse profile variations,  and torque changes, might be instrumental for our understanding of the processes at the basis of radio emission in pulsars, the reason for the existence of such metastable states, and what causes the transitions between them, are still unknown.  The fact that radio emission is of little relevance in the overall energy budget of rotation-powered pulsars might suggest that only some minor local effect in the emission region is involved.  On the other hand, recent findings seem to imply the occurrence of global reconfigurations of the whole pulsar magnetosphere and system of currents. Such findings include the discovery of  variable spin-down rate in intermittent pulsars \cite{kra06}, the observation of timing noise patterns correlated with pulse profile changes \cite{lyn10}, and the detection of nearly simultaneous mode-switching on both magnetic poles \cite{wel12}.

Can X-ray observations say something in this respect? Global magnetospheric changes were invoked to interpret the variability of \psr .
In particular, the  changes in the thermal component were attributed to varying scattering and/or absorption related to an expansion of the volume of the closed magnetosphere  \cite{her13}.  In principle, a similar scenario might also account for the more complex situation derived from the new data.
This  would imply that we are    only {\it partially} seeing the thermal flux emitted from a hot spot on the star surface. The real luminosity would then be higher than the observed one.  However, we note that such explanations based on absorption probably require an unrealistically large particle density in order to reach a sufficient optical thickness \cite{mer13}.

More local origins of the bimodal X-ray properties, limited to the the regions above the polar caps,  can   be invoked in the context of the PSG model (Sect.~\ref{sec-psg}). The mode switching could reflect two states in which the dominant inverse Compton scattering occurs either at the resonant frequency or at the peak of the  distribution of thermal photons from the surface   \cite{zha97}.

It is also interesting to estimate the amount of energy involved in the mode transitions, based on  the observed change in X-ray luminosity. The bolometric luminosity of the thermal component (assuming a blackbody spectrum) changes by a factor 2.7. This corresponds to an increase of $\sim$6 $\times$ 10$^{28}$ erg s$^{-1}$ during the Q-mode. This amounts only to   0.06\%   of the rotational energy loss of the pulsar.  The non-thermal flux in the 0.5-2 keV range doubles from the B- to the Q-mode, and the additional luminosity is $\sim$0.15\% of the rotational energy loss.

\subsection{Comparison with other mode-switching pulsars}

Unfortunately, most mode-switching radio pulsars are too faint for detailed studies in the X-ray range with the current facilities.
The only other object of this class for which simultaneous radio/X-ray observations have been carried out is PSR B1822--09. Its  timing parameters  are $P$ = 0.77  s  and  $\pdot$ = 5.3 $\times$ 10$^{-14}$ s s$^{-1}$.  It is therefore younger ($\tau$ = 0.2 Myr) and more energetic ($\dot{E}_{rot}$ = 4.6 $\times$ 10$^{33}$ erg s$^{-1}$) than \psr . 
Its radio emission is characterized by  a main pulse and an interpulse separated by about 0.5 in phase \cite{bac10,lat12}. The pulse and interpulse behave in anticorrelated manner when the pulsar switches between a  Q- and a B-mode. The mode changes occur much more frequently than in \psr : the modes in PSR B1822--09  have average durations of only a few minutes. 
Given that PSR B1822--09 is most likely a nearly orthogonal rotator,  it is particularly interesting to compare its   X-ray properties  with those of \psr\ in order to understand if the  different geometry plays an important role.

Long observations with \xmm\  (200 ks of effective exposure) and several radio telescopes (WSRT, GMRT, Lovell) were carried out in 2013--2014  \cite{her17}.
In both modes the X-ray spectrum was well described by the sum of two blackbodies with temperatures $kT_1$ = 0.08 keV,  $kT_2$ = 0.19 keV  and emitting radii  $R_1$ = 2 km, $R_2$ = 100 m.   
X-ray pulsations were detected during both radio modes and could be ascribed to the hotter of the two blackbody components \cite{her17}.  In this respect, it is interesting to note that the total spectrum of \psr\ during the Q-mode could also be fitted with the sum of two blackbodies \cite{mer16}, but  the pulse-resolved spectral analysis favoured a power-law rather than a cooler blackbody for the unpulsed component.

At variance with what found in \psr\ no variations correlated to the radio modes were found in  the X-ray emission of PSR B1822--09 \cite{her17}. This could be due to the different geometry, to the short duration of the modes in PSR B1822--09, or simply to the lack of adequate sensitivity to reveal small differences in the X-ray properties. 

Finally,  an X-ray source with a harder spectrum and no bright optical counterpart was discovered at 5$''$ from the pulsar position. If this hard emission is due to a pulsar wind nebula, this would be another difference with respect to \psr , possibly related to the higher $\dot{E}_{rot}$ of PSR B1822--09.

 \section{Conclusions}
 
 Extensive radio observations of the  prototypical  mode-switching   pulsar \psr , carried out for more than 40 years, have provided a wealth of information and a detailed knowledge of its complex phenomenology. Thanks to the high sensitivity of the \xmm\ satellite, it has recently become possible to complement these studies with a high-energy view.
 
It turned out that \psr\ exhibits a very interesting behaviour  also in the X-ray band,  being the first rotation-powered pulsar for which  X-ray variability  (not related to magnetar-like activity as, e.g., in PSR J1846--0258 \cite{gav08})  has been detected.  Studying how the X-ray properties (flux, spectrum, pulsed fraction) vary in connection with the pulsar radio-mode can provide important information on the physical conditions in the acceleration regions and on the processes responsible for the pulsar multiwavelength emission.  
 
Furthermore, the results reviewed in the previous sections, motivate deep searches for X-ray variability phenomena in other radio pulsars -- a research line that received little attention up to now. Studies of pulsars showing different radio states will greatly benefit of the  advent of future X-ray missions with a large collecting area in the soft X-ray band. These might also provide X-ray data allowing to discriminate between different states even in the lack of simultaneous radio data.








\section*{Acknowledgement}

SM is grateful to all the colleagues who collaborated to the study of \psr\ for stimulating discussions, important contributions to the data analysis, and support in the  organization of a complex observational campaign. We acknowledge the useful comments of an anonymous referee.

\balance


 

\end{document}